\documentclass[5p,twocolumn]{elsarticle}

\usepackage[T1]{fontenc}
\usepackage[utf8]{inputenc}
\usepackage{amsmath}
\usepackage{newtxtext,newtxmath}
\usepackage{microtype}
\usepackage{textcomp}
\usepackage{graphicx}
\usepackage{booktabs}
\usepackage{array}
\usepackage{tabularx}
\usepackage{xcolor}
\usepackage{placeins}
\usepackage{xurl}
\graphicspath{{figures/}}
\usepackage[hidelinks]{hyperref}
\biboptions{sort&compress}

\renewcommand{\topfraction}{0.9}
\renewcommand{\bottomfraction}{0.8}
\renewcommand{\textfraction}{0.07}
\renewcommand{\floatpagefraction}{0.7}

\newcommand{\Fven}{\textit{F.~venenatum}}
\newcommand{\Aniger}{\textit{A.~niger}}
\newcommand{\Oxy}{\texorpdfstring{O\textsubscript{2}}{O2}}
\newcommand{\flux}{mmol\,gDW$^{-1}$\,h$^{-1}$}
\newcommand{\gl}{g\,L$^{-1}$}
\newcommand{\glcfrukey}{Sugar concentrations in g\,L$^{-1}$ (Glc glucose, Fru fructose).}

\journal{Biochemical Engineering Journal}

\begin{document}

\begin{frontmatter}

\title{Rule-based curation and validation in batch fermentation of a genome-scale
metabolic model of the mycoprotein fungus \texorpdfstring{\textit{Fusarium venenatum}}{Fusarium venenatum} A3/5}

\author{Jaewook Lee}
\author{Thomas Vinestock}
\author{Miao Guo\corref{cor1}}
\cortext[cor1]{Corresponding author.}
\ead{miao.guo@kcl.ac.uk}
\affiliation{organization={Department of Engineering, Faculty of Natural, Mathematical \& Engineering Sciences, King's College London},
  addressline={Strand}, city={London}, postcode={WC2R 2LS}, country={United Kingdom}}

\begin{abstract}
Sustainable protein supply is a central food-system challenge, and mycoprotein from \textit{Fusarium venenatum} is an established, low-impact option.
Its production relies on refined glucose, and yields on lower-cost feedstocks are normally found by fermentation.
Genome-scale metabolic models (GEMs) can estimate these yields before any experiment, but to our knowledge none has been published for this fungus, and models transferred from related species inherit template errors.
Here we built iJL1143, a GEM of strain A3/5, from an \textit{Aspergillus niger} model and curated it with rules based on homology, thermodynamics and the literature.
We compared it with the uncurated Draft model in four oxygen-enriched batch fermentations on glucose and fructose.
Thermodynamic directionality rules and the direction of the mitochondrial GTP/GDP carrier brought the maximum ATP yield of the Draft model within the theoretical range.
As a result, with sequential sugar uptake and no fitted parameters, the mean final-biomass discrepancy fell from 20\% for the Draft model to 6\% for iJL1143.
Sequential uptake and a fitted lag time also reduced its biomass time-course error from 4.88 to 0.42~\gl{}.
In oxygen-enriched batch fermentation, iJL1143 can therefore estimate obtainable biomass before fermentation, which could reduce the experiments needed to develop lower-cost mycoprotein.
\end{abstract}


\end{frontmatter}

\section{Introduction}

Mycoprotein produced by aerobic fermentation of \textit{Fusarium venenatum} A3/5 is an established food protein with lower greenhouse gas emissions and land use than many livestock-derived proteins \citep{finnigan2025,upcraft2021,wiebe2002}.
Industrial production relies primarily on refined glucose \citep{vinestock2026}.
The biomass formed per unit of sugar influences both the choice of feedstock and the cost of the process.
Glucose and fructose mixtures from sucrose hydrolysates and from food and beverage side-streams offer candidate low-cost feedstocks \citep{vinestock2026}.
A model that estimates biomass yield and the course of a fermentation on such feedstocks could reduce the experiments needed to assess them.

Kinetic models describe such fermentations closely, but they take their yields from the data they fit.
A dual-substrate Monod model of glucose and fructose fermentations used 15 fitted parameters, two of them yield coefficients \citep{vinestock2026}.
A yield fitted in this way cannot be estimated before the experiment.

Genome-scale metabolic models (GEMs) instead estimate yield from the stoichiometry and energetics of the metabolic network \citep{thiele2010,orth2010}.
Dynamic flux balance analysis (dFBA) solves a linear programme at each time step and uses the resulting fluxes to update biomass and extracellular concentrations through mass balances \citep{mahadevan2002}.

GEMs exist for industrial filamentous fungi, including \textit{Aspergillus niger} \citep{brandl2018} and \textit{Fusarium fujikuroi} \citep{li2023}.
To our knowledge, no GEM of \Fven{} has been published. Transferring a network from another related species may propagate reaction assignments or directionality constraints that are unsupported in the target organism.
One consequence of inconsistent reaction directionality is the formation of thermodynamically infeasible energy-generating cycles, and they occur in over 85\% of models without extensive manual curation \citep{fritzemeier2017}.
Curation can remove these errors, but its steps are often hard to reproduce or to transfer to another species.

Rules based on homology, thermodynamics and the literature offer a reproducible alternative.
Comparing the curated model with the uncurated draft then shows what the rules contribute.
Oxygen-enriched batch fermentation offers a clean first test, because a growth-maximising model is most defensible when oxygen does not limit growth.

We therefore hypothesised that, in oxygen-enriched batch fermentation, rule-based curation alone would bring predicted final biomass closer to measurement, because yield is set by network stoichiometry and energetics.
Time courses also depend on sequential sugar uptake and on the lag phase, which a growth-maximising model does not contain.
We expected that adding both would substantially improve agreement with the measured time courses.

Here we reconstruct iJL1143, a GEM of \Fven{} A3/5, from the \Aniger{} model iJB1325 and curate it with rules based on homology, thermodynamics and the literature.
We compare the rule-curated model with the uncurated Draft model across four initial glucose--fructose compositions under oxygen-enriched batch fermentation conditions.
We also add sequential sugar uptake and lag time, examine measured by-products that the model fails to reproduce, and test whether medium pH can be explained by growth-associated stoichiometry.

\section{Materials and methods}

\subsection{Model reconstruction and curation}\label{sec:curation}

\subsubsection{Template model and orthology mapping}\label{sec:orthology}

Template-based reconstruction builds a draft model for a new species by transferring the reactions of an existing model whose genes have homologues in the new species \citep{wang2018}.
The transfer therefore depends on how genes are paired between the two genomes.
Reciprocal best hits, pairs of genes in two genomes that are each other's best hit, appear to be the most common working definition of orthology \citep{morenohagelsieb2008}.
Orthogroups extend orthology to several species as sets of genes descended from a single gene in their last common ancestor \citep{emms2019}, and so also capture the one-to-many relationships that reciprocal best hits cannot.
The paired genes enter the model through gene--protein--reaction associations, called gene associations here, which connect genes, proteins and reactions through logical AND and OR relationships \citep{thiele2010}.

The CBS 513.88 version of iJB1325 (2,320 reactions, 1,373 genes) served as the template \citep{brandl2018}.
We chose it because it is a community-curated model of an industrial filamentous fungus that passed 373 of 471 tests of growth, gene deletion and network function \citep{brandl2018}.
Its genes were mapped to the \Fven{} A3/5 proteome \citep{king2018} by reciprocal best hits from blastp in BLAST+ 2.16.0 \citep{camacho2009}, with at least 40\% identity, at least 70\% query coverage and $E \leq 10^{-30}$.
These cut-offs follow those used to build the \textit{Penicillium chrysogenum} model with RAVEN, except for a coverage of 70\% instead of 50\% and no minimum alignment length \citep{agren2013}.
To check these pairs and to find orthologs that a one-to-one search misses, orthogroups were also built across 10 fungal proteomes with OrthoFinder 2.5.5, with DIAMOND 2.2.8 as the sequence aligner \citep{emms2019,buchfink2021}.
For the Curated model, template genes without a reciprocal best hit were then mapped to the \Fven{} members of their orthogroup.
Gene associations were rewritten as Boolean rules over \Fven{} genes, and a complex was retained when at least one of its subunits had an ortholog.
Nine subunits of complexes, in nine reactions, had no ortholog.
To search for a homologue that the reciprocal search had missed, each was aligned against the whole proteome by Smith--Waterman local alignment \citep{smith1981}.
The alignment used the BLOSUM62 matrix, gap penalties of 11 for opening and 1 for extension, and the same identity and coverage thresholds.
Only the sulfur carrier of molybdopterin synthase matched a protein of the same function, which took its place in the association, so the other associations name only the subunits that have orthologs.
Finally, to assess the evidence behind each association, the Enzyme Commission (EC) number of the reaction was compared with the UniProt EC annotation of its gene \citep{uniprot2025}.

\subsubsection{Rule-based curation}\label{sec:rules}

Curation converts a draft reconstruction into a curated one by re-evaluating its content \citep{thiele2010}.
It adds reactions missing from the network by gap-filling and defines the exchange reactions, which represent the system boundary through which metabolites enter and leave \citep{thiele2010}.
Curation also sets the direction of each reaction, and thermodynamics provides a measure for this in the reversibility index, $\Gamma$, the range of metabolite concentrations required to reverse a reaction \citep{noor2012}.
The further $\Gamma$ is from 1, the more irreversible the reaction.
A value of 1,000 means that reversal would need concentrations spanning three orders of magnitude, about 3~\textmu M to 3~mM around 100~\textmu M \citep{noor2012}.
eQuilibrator estimates the standard Gibbs energies behind the index with the component contribution method \citep{noor2013,beber2022}.
Finally, curation sets the energy cost of maintenance, represented as ATP hydrolysis, growth-associated within the biomass reaction and non-growth-associated in a reaction of its own \citep{thiele2010}.
To compare predicted growth with observed yields, biomass is defined to have a molecular weight of 1~g\,mmol$^{-1}$ \citep{chan2017}.

Six stages of rules were applied to obtain iJL1143, the Curated model (Fig.~\ref{fig:reconstruction}a, Supplementary Table~S1).
Stage~1 retained only reactions supported by orthology, which removed 85 template reactions.
Because orthology can miss a gene whose function is annotated, Stage~2 restored six reactions supported by the UniProt EC annotation of the proteome, and gap-filling added three reactions needed for biomass synthesis.
Stage~3 required evidence for secretion.
A metabolite could leave only through a gene-supported transporter or with literature support, which closed 330 exchange reactions.
Stage~4 applied redox rules that remove cost-free routes between NADH and NADPH and between the cytosolic and mitochondrial pools, because such routes would let the model convert one cofactor into the other without an energy cost.
It also made NAD(P)-dependent reactions irreversible where $|\ln\Gamma| \geq 7$, that is, $\Gamma$ of about 1,100 or more, and the estimate lay more than two standard errors from zero, unless that direction blocked biomass synthesis.
$\Gamma$ was computed for concentrations around 1~mM at the default conditions of eQuilibrator, pH 7.5, pMg 3, ionic strength 0.25~M and 25\,\textdegree C.
These conditions were used in every compartment, and the rule was not applied to transport reactions.
Because biomass yield depends on what the biomass is made of, Stage~5 normalised biomass at 1~g\,mmol$^{-1}$ to the macromolecular composition of growing A3/5 mycelium (Supplementary Table~S2).
Protein (37\% of dry mass) and RNA (9\%) were taken from measurements in this strain before RNA reduction \citep{towersey1977,trinci1992}.
Cell wall (17\%) and lipid (9\%) were derived from the composition of mycoprotein, the RNA-reduced product of the same strain \citep{fda2002grn91}, and from the 0.68~g of product left per gram of mycelium after RNA reduction \citep{towersey1977}.
DNA (0.9\%) was taken from \textit{Aspergillus oryzae} \citep{pedersen1999}, and the dry mass left after these fractions and 5\% ash \citep{solomons1976,fda2002grn91} was assigned to the small-molecule pool.
Stage~5 also set the maintenance costs, both taken from \Aniger{}, with a non-growth-associated maintenance (NGAM) of 1.9~mmol ATP per gram dry weight (gDW) per hour \citep{andersen2008}.
Growth-associated maintenance (61~mmol ATP per gDW) was taken from the template, which inherits it from iMA871.
There it was calculated from the biomass yield of \Aniger{} on glucose with P/O ratios of 2.64 and 1.64 \citep{andersen2008}.
Stage~6 extended the thermodynamic criterion of Stage~4 to all metabolic reactions.
It also charged the export of organic acids one ATP per negative charge, which extends the ATP cost of lactate export in yeast to divalent anions that leave with an additional proton \citep{vanmaris2004,vanmaris2004aem,abbott2009}.
Finally, it restricted the mitochondrial GTP/GDP carrier to GTP import, its probable physiological direction in \textit{Saccharomyces cerevisiae} \citep{vozza2004}, so that mitochondrial ATP cannot reach the cytosol as GTP without the proton cost of the ADP/ATP carrier.
iJL1143 contains all six stages and no other change.
The variants in which stages were removed or added, and the inputs changed in Section~\ref{sec:energy}, were analysed only to explain its behaviour and are not part of the model.

Both models retain the respiratory stoichiometry of the template, which determines their P/O ratios.
The P/O ratio is the number of ATP formed per oxygen atom reduced.
In the template stoichiometry, each ATP delivered to the cytosol uses 4.88 protons, 3.88 at ATP synthase and one at the ADP/ATP carrier.
With 12 protons pumped per NADH and 8 per succinate, this results in P/O ratios of 2.46 and 1.64.
This stoichiometry was retained because these ratios lie close to the measured mitochondrial ratios of about 2.5 and 1.5 \citep{hinkle2005} and to the 2.64 and 1.64 set in iMA871 \citep{andersen2008}, the source of the growth-associated maintenance.
Because these ratios are taken from the template and were not measured in \Fven{}, lower P/O ratios were tested in the sensitivity analysis (Section~\ref{sec:energy}), but iJL1143 retains the template values.

\subsubsection{Draft model}\label{sec:draft}

The Draft model is the comparator for every test (Fig.~\ref{fig:reconstruction}a).
It represents the uncurated starting point of a reconstruction, which can contain reactions that curation later removes \citep{thiele2010}.
The Draft model therefore retains every reaction of the template model, with gene associations rewritten through reciprocal best hits only.
To isolate the effect of the network, it takes the biomass composition and maintenance of the Curated model, with non-growth-associated maintenance as one added reaction, so the two models differ in their network and gene associations and not in biomass or maintenance.
A draft built with an orthology filter would lack 85 template reactions, nine of which Stage~2 restores.
With these nine retained, this filter changes neither the maximum ATP yield nor the biomass yield of the Draft model (Supplementary Table~S3), so the choice of draft does not affect the yield comparison.

\subsection{Simulation and analysis}

\subsubsection{Dynamic flux balance analysis}\label{sec:dfba}

Flux balance analysis (FBA) predicts the flow of metabolites through a metabolic network, and from it the growth rate, without kinetic parameters \citep{orth2010}.
It assumes that internal metabolites are at steady state, so the stoichiometric matrix $\mathbf{S}$ and the flux vector $\mathbf{v}$ satisfy a mass balance, and it sets lower and upper bounds on each flux.
These constraints allow a range of flux distributions, and linear programming selects the one that maximises an objective, which for growth is the biomass flux $v_\mathrm{bio}$ (Eq.~\ref{eq:fba}).
\begin{equation}
  \begin{aligned}
    \max_{\mathbf{v}}\quad & v_\mathrm{bio}\\
    \text{subject to}\quad & \mathbf{S}\mathbf{v}=\mathbf{0},\quad \mathbf{v}_\mathrm{min}\leq\mathbf{v}\leq\mathbf{v}_\mathrm{max}
  \end{aligned}
  \label{eq:fba}
\end{equation}
FBA itself does not predict metabolite concentrations and, in its basic form, does not account for regulation.
dFBA adds the time course by solving Eq.~\ref{eq:fba} at each time step and then updating the extracellular concentrations \citep{mahadevan2002}.

dFBA simulated the four batch fermentations (Section~\ref{sec:cultures}) by maximising growth within sugar uptake and oxygen bounds.
At each time step, COBRApy \citep{ebrahim2013} with the GLPK solver solved Eq.~\ref{eq:fba} with sugar uptake bounds from Michaelis--Menten kinetics, and biomass and extracellular concentrations were then advanced by a carbon-conserving explicit Euler step.
A step of 0.5~h was sufficient for these comparisons.
Reducing it to 0.1~h changed the mean discrepancy in final biomass by less than 0.3 percentage points for either model and, with lag time refitted, the biomass RMSE by less than 0.01~\gl{}.
Runs covered 150~h. Steps without a feasible solution occurred only after all sugars were depleted, when maintenance could not be met, and were recorded with zero flux, so that biomass then stayed constant.
Because FBA does not predict how fast sugars are taken up, the maximum uptake rate of each model was set from the growth rate of the industrial process.
It equals 0.20~h$^{-1}$, the upper end of that growth rate \citep{wiebe2002}, divided by the aerobic biomass yield of the model at an uptake of 1~\flux{}.
Because maintenance then accounts for a smaller fraction of the ATP demand, growth at this uptake is 0.20--0.21~h$^{-1}$.
Growth rate is therefore an input of the simulations and not a prediction, whereas biomass yield follows from the network.
A $K_\mathrm{m}$ of 0.26~mM was assumed for both sugars, between the values reported for the \Aniger{} transporter MstA on glucose (0.025~mM) and fructose (4.5~mM) \citep{vankuyk2004}.
Its effect was tested across that range (Section~\ref{sec:energy}).
Specific \Oxy{} uptake was bounded by the maximum oxygen transfer rate divided by the biomass concentration, so that oxygen limits growth only when the demand of the culture exceeds transfer.
All inputs and their origins are listed in Table~\ref{tab:inputs}.

Runs were compared in four cumulative settings that add sequential sugar uptake and then lag time.
These settings change only the simulation and not the metabolic network of either model.
The base setting used an initial biomass of 0.01~\gl{} and simultaneous sugar uptake, with each sugar at its own maximum uptake rate.
Filamentous fungi, however, use a readily available carbon source in preference to others through carbon catabolite repression \citep{adnan2018}.
The second setting therefore represented sequential uptake, which blocks fructose uptake while glucose remains above 0.5~mM.
This threshold is assumed, so thresholds of 0.05 and 5~mM were also tested, with lag time refitted.
They changed the discrepancy in final biomass of the mixed fermentations by less than 0.3 percentage points and their biomass RMSE by 0.1~\gl{} or less, so the choice of threshold does not affect the comparison.

The third and fourth settings represented the lag phase by fitting the initial biomass, first, as one shared value and then, as one value per condition.
This approach is valid because a lower initial biomass delays growth but leaves final biomass nearly unchanged.
Under exponential growth it is equivalent to a lag time, because the product of lag time and growth rate is a transformation of the initial physiological state of the inoculum \citep{baranyi1994}.
The fitted value therefore combines inoculum size and lag.
The fit matched the simulated time to half final biomass to the measured one, at which the mean online biomass of the replicates reaches half the mean final dry weight.
Values per condition were obtained by a root search between $10^{-6}$ and 0.03~\gl{}, and the shared value minimised the summed squared timing error over the four conditions (Supplementary Note~1, Supplementary Fig.~S1).
Each model was fitted separately, because the two models reach different final biomass at different maximum uptake rates, so an initial biomass fitted to one model would not align the other.
A value per condition is justified because each fermentation used its own inoculum, and our previous work suggested that differences in lag reflect the age and viability of the conidia \citep{vinestock2026}.

\begin{table*}[tp]
  \centering
  \caption[Model and fermentation inputs and their origin]{\textbf{Model and fermentation inputs and their origin.}}
  \label{tab:inputs}
  \small
  \begin{tabularx}{\textwidth}{@{}>{\raggedright\arraybackslash}p{3.4cm} >{\raggedright\arraybackslash}X >{\raggedright\arraybackslash}p{2.3cm} >{\raggedright\arraybackslash}X@{}}
    \toprule
    Input & Value & Origin & Source \\
    \midrule
    Template model & iJB1325 (\Aniger), CBS 513.88 version & literature & \citep{brandl2018}, Additional file 3 \\
    Biomass composition & Fractions of growing mycelium (Supplementary Table~S2), 1~g\,mmol$^{-1}$ & measured (protein, RNA), derived (others) & \citep{towersey1977,trinci1992,fda2002grn91,pedersen1999} \\
    Non-growth-associated maintenance & 1.9~mmol ATP gDW$^{-1}$\,h$^{-1}$ & literature (\Aniger) & \citep{andersen2008} \\
    Growth-associated maintenance & 61~mmol ATP gDW$^{-1}$ & literature (\Aniger) & \citep{andersen2008} \\
P/O ratio & 2.46 for NADH, 1.64 for succinate & literature (template) & 12 and 8 protons pumped, 4.88 per cytosolic ATP \\
    Acid export cost & 1 ATP per negative charge at pH 7 & literature & \citep{vanmaris2004,vanmaris2004aem,abbott2009} \\
    Growth rate that sets the maximum sugar uptake & 0.20~h$^{-1}$ & literature & growth rate of the Quorn process, 0.17--0.20~h$^{-1}$, \citep{wiebe2002} \\
    Sugar $K_\mathrm{m}$ & 0.26~mM & assumed & between the glucose (0.025~mM) and fructose (4.5~mM) values of \Aniger{} MstA, \citep{vankuyk2004} \\
    Maximum oxygen transfer rate & 46.9~mmol\,L$^{-1}$\,h$^{-1}$ & measured ($k_\mathrm{L}a$) & $k_\mathrm{L}a$ 96~h$^{-1}$, 200\% air \citep{vinestock2026} \\
    Initial biomass & 0.01~g\,L$^{-1}$ in the base settings; fitted per condition & assumed; fitted & time to half final biomass, which sets the lag time \\
    Sequential sugar uptake & other sugars blocked above 0.5~mM glucose & assumed & carbon catabolite repression, \citep{adnan2018} \\
    Initial sugars & 15~g\,L$^{-1}$ in total, as prepared & measured (recipe) & \citep{vinestock2026} \\
    Medium & KH$_2$PO$_4$ 203~mM, NH$_4$Cl 76~mM in the broth (46~mL of medium per 50~mL) & measured (recipe) & \citep{vinestock2026} (KH$_2$PO$_4$ 30~g\,L$^{-1}$, volumes), \citep{whittaker2022} (NH$_4$Cl 4.4~g\,L$^{-1}$) \\
    Ionic strength & 0.446~M, phosphate p$K_{\mathrm{a}2}'$ 6.79 & measured (recipe) & calculated with the Davies equation and the charge balance \\
    \bottomrule
  \end{tabularx}

  \smallskip
  \parbox{\textwidth}{\footnotesize Only initial biomass, which sets the lag time, is fitted. Final biomass barely depends on it.}
\end{table*}

\subsubsection{Energy tests, stage removal, sensitivity and oxygen limitation}\label{sec:energy}

To test whether the energetics of each model stay within physical limits, its maximum ATP yield, the largest amount of ATP it can form per mol glucose, was computed on aerobic minimal medium with no maintenance.
Energy-generating cycles were tested by asking whether a cofactor can be recharged with no substrate uptake \citep{fritzemeier2017}, for nucleoside triphosphates, reduced cofactors and the mitochondrial proton gradient.
Two further cycle tests asked whether NADH can be converted to NADPH at no cost, in the cytosol and in mitochondria, which are the routes that the redox rules of Stage~4 target.
To find the effect of each curation stage, the reactions listed under that stage in the change log were returned to their Draft state in the Curated model.
Separately, the same reactions were given their Curated state in the Draft model (Supplementary Table~S3).
Because the two sets of directionality rules might substitute for each other, they were also removed together from the Curated model.
In the Draft model, the carrier rule was combined with each set and with both, and these models were run in the batch fermentations like the Draft and Curated models.

Several inputs are borrowed or derived rather than measured in this strain.
The sensitivity of the yield to each was therefore computed at the maximum sugar uptake rate of the unchanged model, with \Oxy{} not limiting (Supplementary Note~2, Supplementary Fig.~S2).
Each macromolecular mass fraction was changed by $\pm$25\%, with the other fractions rescaled so that biomass stays 1~g\,mmol$^{-1}$ \citep{choi2023}, and each ATP cost was changed by the same amount.
The P/O ratio for NADH was reduced to 2.25 and 2.0 by increasing the number of protons used at ATP synthase.
The assumed $K_\mathrm{m}$ was varied from 0.025 to 4.5~mM with initial biomass held at its fitted value.
Model quality was scored with MEMOTE 0.17, a standardised set of tests for genome-scale metabolic models \citep{lieven2020} (Supplementary Table~S4).

Oxygen limitation was tested as an explanation of glycerol formation, both by comparison with the experimental data and in the models.
In the experimental data, glycerol concentrations were compared with the dissolved \Oxy{} and online biomass readings at each high-performance liquid chromatography (HPLC) sample.
The timing of events used the nearest reading, whereas rates and comparisons between replicates used readings smoothed by a running median over five points and interpolated to the sample time.
The net glycerol rate between two samples is the change in concentration divided by the integral of online biomass over that interval, and its relation to dissolved \Oxy{} is given as a Pearson correlation coefficient.
In both models, specific \Oxy{} uptake was capped at 80, 60, 40 and 20\% of the smallest \Oxy{} uptake that allows maximal growth at the maximum sugar uptake rate, with sugar uptake either bounded, as in dFBA, or fixed at that rate.
Flux variability analysis, which finds the range of each flux across solutions with the same objective value \citep{mahadevan2003}, was then used to obtain the export ranges at the growth optimum.
A further test of the Curated model asked whether an \Oxy{} shortfall could produce glycerol at the measured rate.
It fixed sugar uptake, capped \Oxy{} uptake below that requirement and left glycerol as the only carbon export, with the inorganic exports that can carry electrons closed.

\subsubsection{Medium pH model}\label{sec:ph}

Medium pH was computed from a charge balance over phosphate, ammonium and the measured organic acids, without fitted parameters.
A charge balance finds the pH at which the charges of all ions in a solution sum to zero, so pH follows once the amount of each ion is known.
Growth changes these amounts in two ways.
One proton is released per ammonium nitrogen assimilated, as measured in yeast chemostats \citep{castrillo1995}, and phosphate and sulfate are taken up according to the biomass composition (6.25~mmol N, 0.38~mmol P and 0.07~mmol S per gDW).
Because the ionic strength of the medium is high, 0.446~M, activity coefficients were taken from the Davies equation for electrolytes in water \citep{davies1938}, which gives an apparent phosphate p$K_{\mathrm{a}2}$ of 6.79.
The balance was driven by the online biomass of each replicate, and the net charge of the ions it does not track was set once from the initial pH of each replicate.
No other quantity was adjusted, and the effects were added in an order fixed beforehand, so that no step could be tuned to the measured pH.

\subsubsection{Error metrics and statistics}\label{sec:metrics}

Final biomass tests the models because neither biomass composition nor maintenance was derived from these fermentations, as recommended for validating growth predictions \citep{kaste2024}.
Its discrepancy is the difference between the predicted value and the measured mean, given in \gl{} and as a percentage of the measured mean.
The absolute value of this difference is plotted, and its mean over the four conditions is the mean discrepancy.
Predicted biomass is ash-free, whereas measured dry weight includes ash, and the two were compared without an ash correction.
The ash of A3/5 mycelium, about 5\% of dry mass (Supplementary Table~S2), is an upper bound for the inorganic mass that the model lacks.
It is only an upper bound because part of it is the phosphorus of nucleic acids and lipids, which the model includes.

Time-course root mean square error (RMSE) was computed over all replicates and measured time points of a condition, with biomass against online biomass and sugars against HPLC, and then averaged over the four conditions.
For comparison with a kinetic model of the same fermentations \citep{vinestock2026}, the mean absolute error (MAE) was computed in the same way.
The pH RMSE was computed per replicate and averaged over the 12 replicates.
Values marked $\pm$ are the standard deviation of three replicates.
No statistical tests are reported, because each condition has three replicates and differences between conditions are not claimed.

\subsection{Batch fermentation data}

\subsubsection{Fermentations and online measurements}\label{sec:cultures}

The batch fermentations on glucose and fructose were reported in our previous work \citep{vinestock2026}.
\Fven{} A3/5 was inoculated at 10,000 spores per mL into 50~mL of Rank Hovis McDougall (RHM) medium in 250~mL baffled shake flasks at 28\,\textdegree C and 220~rpm.
Total sugar was 15~\gl{} in four conditions that differ in the ratio of glucose to fructose, written Glc~15, Glc~10 + Fru~5, Glc~5 + Fru~10 and Fru~15 (glucose and fructose in \gl{}), with three replicates each.
To increase oxygen transfer, the humidified headspace contained 200\% of the oxygen of air, and the volumetric mass transfer coefficient, $k_\mathrm{L}a$, was 96~h$^{-1}$.
With twice the air-saturation \Oxy{} concentration of water at 28\,\textdegree C, this corresponds to a maximum oxygen transfer rate of 46.9~mmol\,L$^{-1}$\,h$^{-1}$, which bounds \Oxy{} uptake in the models (Section~\ref{sec:dfba}).
Online biomass, pH and dissolved \Oxy{} were recorded with PreSens shake flask readers.
Online biomass is inferred from reflectance and was calibrated to the start and end points of each replicate.
Online readings and HPLC samples are placed on a common time axis, which starts at the first HPLC sample.

\subsubsection{Final biomass and yield}\label{sec:yield}

Yield is the gravimetric final biomass concentration divided by the nominal initial sugar concentration.
Final biomass was measured by vacuum filtration and drying to constant mass at 65\,\textdegree C.
Its concentration is the dry weight divided by the effective final volume, which is the initial volume minus all sampled volumes (pH, HPLC and biomass).
Biomass and sugars do not evaporate, so water loss does not affect this value.
One Fru~15 replicate lost 18.6~mL more volume than was removed by sampling, compared with 3.6 and 6.6~mL for the other two.
This replicate was included, and excluding it would change the Fru~15 yield from 0.53 to 0.54~g\,g$^{-1}$.

\subsubsection{Liquid chromatography and peak identity}\label{sec:hplc}

Sugars and products were measured by HPLC on a Bio-Rad Aminex HPX-87H column with 0.001~M H$_2$SO$_4$ at 0.4~mL\,min$^{-1}$ and 30\,\textdegree C.
The injection volume was 5~\textmu L for Glc~15 and Glc~5 + Fru~10 and 10~\textmu L for the other two, and a Shimadzu RID-20A detector recorded refractive index.
Absorbance from 190 to 254~nm was recorded by a photodiode array detector placed before it and was used only to distinguish carboxylic acids from polyols in peak identification (Supplementary Note~3).
Sugar concentrations are peak areas scaled to the nominal initial concentration of each run, and glycerol, ethanol and acetate were quantified with external calibration factors.
Peak identity was checked in the raw instrument files against single-analyte standards (Supplementary Note~3).
Medium phosphate elutes at 11.5--12.4~min and covers pyruvic and citric acids, so these two could not be measured.

A peak at 16.1~min, tentatively assigned to arabitol (Supplementary Note~3, Supplementary Fig.~S3), lies on the tail of the fructose peak, so this tail was removed before integration.
To remove it, a chromatogram of the same flask taken before any product formed was scaled and shifted to match the tail between 15.2 and 15.6~min and then subtracted.
The peak was then integrated from 15.6 to 16.9~min above a straight baseline.
Without fructose, the resulting area matches the peak table of the instrument software, with a median ratio of 1.02.
On Fru~15, without product, it was $-165$ to $+31$ per \textmu L, compared with maxima of 520--2,100 per \textmu L in the fermentations containing glucose, so the remaining background is small against the peak.

\section{Results}

We compared two models in four batch fermentations of \Fven{} A3/5 with 15~\gl{} total sugar, using two tests.
The Draft model retains every template reaction, and the Curated model, iJL1143, applies six stages of rules to it (Fig.~\ref{fig:reconstruction}a).
Final biomass tests the biomass yield of each model with no fitted parameter.
Time-course RMSE then tests how closely the models follow the fermentations once sequential uptake and lag time are added (Section~\ref{sec:metrics}).
We first describe what curation changed in the network and its energetics, then test both models against the fermentations, and finally examine the products and the medium pH.

\subsection{Model reconstruction and curation}

\subsubsection{Orthology supports most template genes}

\begin{figure*}[tp]
  \centering
  \includegraphics[width=\textwidth,height=0.6\textheight,keepaspectratio]{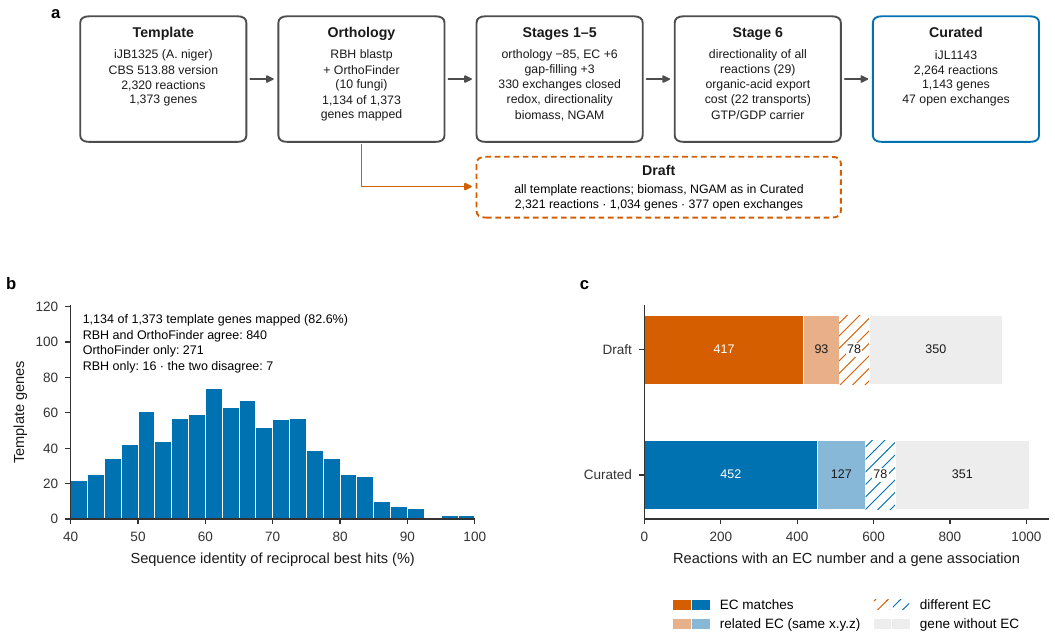}
  \caption[Reconstruction of the \Fven{} A3/5 model]{\textbf{Reconstruction of the \Fven{} A3/5 model and the evidence behind its gene associations.}
  (a) Workflow from the template model to the Draft and Curated models (Section~\ref{sec:curation}, Supplementary Table~S1). 
  (b) Sequence identity of reciprocal best hit (RBH) ortholog pairs, with the agreement between RBH and orthogroups.
  (c) Reactions that carry an EC number and a gene association, by whether the UniProt EC of the associated \Fven{} gene matches the reaction EC, shares its first three levels, differs, or is missing.}
  \label{fig:reconstruction}
\end{figure*}

Orthology supported most template genes, and curation closed most exchange reactions open for secretion.
The model was built from the CBS 513.88 version of the \Aniger{} model iJB1325 (Fig.~\ref{fig:reconstruction}a), and orthology supported 82.6\% of its 1,373 genes.
Reciprocal best hits and orthogroups agreed on 840 of these 1,134 supported genes (Fig.~\ref{fig:reconstruction}b).
The Curated model has 2,264 reactions, 1,797 metabolites and 1,143 genes, 57 fewer reactions and 109 more genes than the Draft model (Table~\ref{tab:models}).
It has more genes despite fewer reactions because its gene associations also draw on orthogroups and on the EC annotation of the proteome, which the Draft model does not use (Sections~\ref{sec:orthology} and \ref{sec:rules}).
Restricting secretion then reduced the exchange reactions open for secretion from 377 to 47.

Curation improved the evidence behind the gene associations only in part, and the MEMOTE score stayed limited.
The number of associations whose gene EC number matches the reaction rose from 417 to 452 (Fig.~\ref{fig:reconstruction}c).
The number with a different EC number, however, stayed at 78, so curation added supported associations without resolving these conflicts.
Both models had a total MEMOTE score of 29\%, the score reported for the template iJB1325 \citep{han2023}, and neither contains gene annotations or Systems Biology Ontology terms (Supplementary Table~S4).

\begin{table}[tbp]
  \centering
  \caption[Properties of the Draft and Curated models]{\textbf{Properties of the Draft and Curated (iJL1143) models.}}
  \label{tab:models}
  \small
  \begin{tabularx}{\columnwidth}{@{}X *{2}{>{\centering\arraybackslash}p{1.45cm}}@{}}
    \toprule
    Property & Draft & Curated \\
    \midrule
    Reactions & 2,321 & 2,264 \\
    Metabolites & 1,818 & 1,797 \\
    Genes & 1,034 & 1,143 \\
    Exchange reactions open for secretion & 377 & 47 \\
    Biomass weight (g\,mmol$^{-1}$) & 1 & 1 \\
    Gene associations whose gene EC matches the reaction EC & 417 & 452 \\
    Gene associations whose gene EC shares only the first three levels & 93 & 127 \\
    Gene associations whose gene EC differs & 78 & 78 \\
    Maximum ATP yield (mol ATP per mol glucose) & 49.1 & 29.9 \\
    Energy-generating cycle tests failed & 2 & 0 \\
    Aerobic biomass yield on glucose (g\,g$^{-1}$) & 0.599 & 0.522 \\
    Aerobic biomass yield on fructose (g\,g$^{-1}$) & 0.599 & 0.523 \\
    \bottomrule
  \end{tabularx}

  \smallskip
  \parbox{\columnwidth}{\footnotesize No experimental data enter this table.}
\end{table}

\subsubsection{Directionality and carrier rules set the maximum ATP yield}

\begin{figure}[tbp]
  \centering
  \includegraphics[width=\columnwidth]{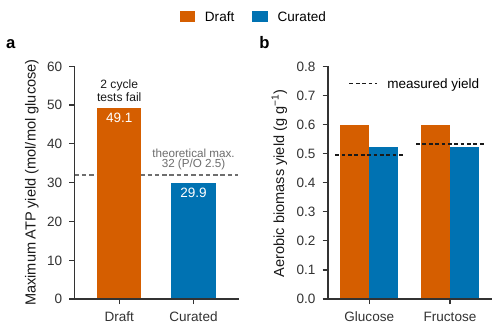}
  \caption[Curation reduces the maximum ATP yield and the biomass yield]{\textbf{Curation reduces the maximum ATP yield and the aerobic biomass yield.}
  (a) Maximum ATP yield per glucose under aerobic conditions, with the number of failed energy-generating cycle tests (Section~\ref{sec:energy}). Dashed line, theoretical maximum of about 32 computed for P/O ratios of 2.5 and 1.5 \citep{hinkle2005}. 
  (b) Aerobic biomass yield at a sugar uptake of 1~\flux. Dashed lines, measured yields in Glc~15 and Fru~15 (Section~\ref{sec:yield}).}
  \label{fig:curation}
\end{figure}

Curation removed the excess ATP yield of the Draft model and reduced the aerobic biomass yield.
The Draft model reached 49.1~mol ATP per mol glucose, above the theoretical maximum of about 32 (Fig.~\ref{fig:curation}a), and failed two energy-generating cycle tests (Section~\ref{sec:energy}).
Both failures were cost-free conversions of NADH to NADPH inherited from the template.
The Curated model instead reached 29.9~mol ATP per mol glucose and failed no test.
The biomass yield fell with the ATP yield.
At a sugar uptake of 1~\flux{}, the aerobic biomass yield on glucose and fructose fell from 0.60~g\,g$^{-1}$ in the Draft model to 0.52~g\,g$^{-1}$ in the Curated model (Fig.~\ref{fig:curation}b).
At the faster maximum uptake used in the fermentations, non-growth-associated maintenance accounts for a smaller fraction of the ATP demand, so the Curated model yields 0.54~g\,g$^{-1}$ on both sugars (Supplementary Note~2).

Reaction directionality and the GTP/GDP carrier rule caused this fall.
To identify the responsible stages, we analysed model variants in which each curation stage was removed from the Curated model or applied alone to the Draft model (Section~\ref{sec:energy}, Supplementary Table~S3).
Two stages set reaction directions from thermodynamics, Stage~4 for NAD(P)-dependent reactions and Stage~6 for all metabolic reactions.
Either set alone reduced the maximum ATP yield of the Draft model to 37.1 and its biomass yield to about 0.56~g\,g$^{-1}$.
Adding the GTP/GDP carrier rule to either set reduced the ATP yield further, to about 30, and the biomass yield to about 0.52~g\,g$^{-1}$, although this rule alone changed neither value.
The variants of the Curated model were consistent with this result.
Removing the carrier rule increased its ATP yield to 37.1, and removing both sets of directionality rules increased the ATP yield to the Draft value of 49.1.
Removing either set of directionality rules alone, or any other stage, changed the maximum ATP yield of the Curated model by at most 0.1 and its biomass yield by at most 0.003~g\,g$^{-1}$.
Added to the Draft model, no other stage changed either value.

Passing the cycle tests did not determine the yield.
A Draft model with the directionality and carrier rules still failed both tests, yet its yield of 0.52~g\,g$^{-1}$ already matched that of the Curated model (Supplementary Table~S3).
Likewise, the Curated model failed both tests at nearly the same yield when the orthology filter, the redox rules or NAD(P) directionality were removed.
Conversely, the Curated model without the carrier rule failed no test at an ATP yield of 37.1, above the theoretical maximum.

Restricting secretion closed 330 exchange reactions with no change in aerobic biomass yield.
Neither removing this stage from the Curated model nor adding it to the Draft model changed either yield (Supplementary Table~S3).

\subsection{Validation in batch fermentation}

\subsubsection{Curation brings predicted final biomass close to measurement}\label{sec:res-biomass}

\begin{figure*}[tp]
  \centering
  \includegraphics[width=\textwidth,height=0.72\textheight,keepaspectratio]{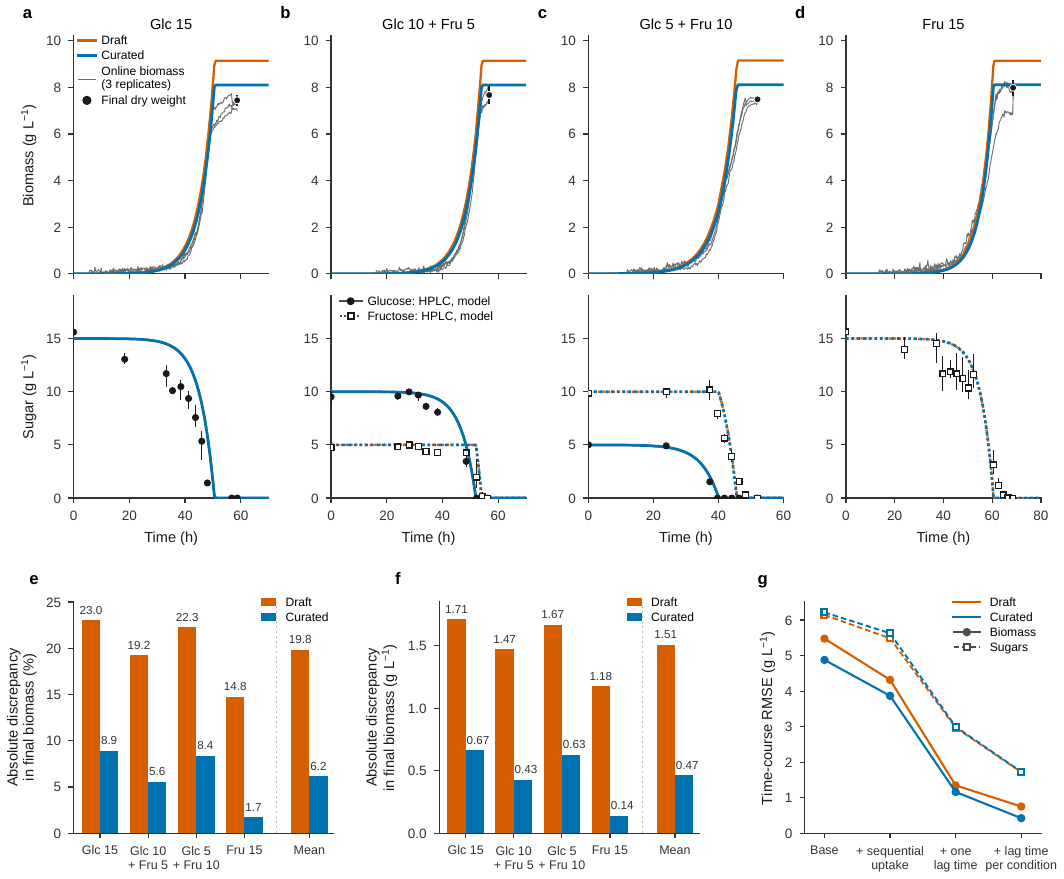}
  \caption[Time courses and final biomass in oxygen-enriched batch fermentation]{\textbf{Curation brings final biomass close to measurement, and sequential uptake and lag time align the time courses.}
  Batch fermentations \citep{vinestock2026} simulated by dFBA with sequential sugar uptake (Section~\ref{sec:dfba}).
  (a--d) One column per condition, with initial biomass fitted per condition to the measured time to half final biomass. Upper panels, biomass: grey lines, online biomass of the three replicates; black point, final dry weight (mean $\pm$ s.d.). Lower panels, sugars: points, HPLC glucose (filled circles) and fructose (open squares), mean and range of the three replicates; lines, simulated glucose (solid) and fructose (dotted).
  (e, f) Absolute discrepancy between predicted and measured final biomass with sequential uptake and no fitted parameters, per condition and as the mean, as a percentage of the measured mean (e) and in \gl{} (f).
  (g) Time-course RMSE over all replicates and measured time points, averaged over the four conditions, as settings are added: base (initial biomass 0.01~\gl{}, simultaneous uptake), sequential uptake, one shared lag time and one lag time per condition.
  \glcfrukey}
  \label{fig:flask}
\end{figure*}

Without fitted parameters, the Curated model predicted final biomass more closely than the Draft model, and the directionality and carrier rules account for the improvement.
Because sugars were depleted in every fermentation, final biomass tests the biomass yield directly.
Measured yields ranged from 0.50 to 0.53~g\,g$^{-1}$ across the four conditions (Supplementary Table~S5).
With sequential uptake, the mean discrepancy was 6\% for the Curated model and 20\% for the Draft model (Fig.~\ref{fig:flask}e), or 0.47 and 1.51~\gl{} (Fig.~\ref{fig:flask}f).
If the full 5\% ash is removed from the measured dry weight, these means rise to 12\% and 26\%, and the Curated model stays closer (Section~\ref{sec:metrics}).
The comparison also does not depend on the uptake rule, because with simultaneous uptake the means were 7\% and 20\%.
The Curated model was closer in all four conditions.
Applied alone to the Draft model, both sets of directionality rules and the carrier rule reduced its mean discrepancy to 6\% (Section~\ref{sec:energy}).
Both this model and the Curated model overpredicted the three fermentations containing glucose by 6--9\% and Fru~15 by 2\%.

This test constrains one yield and does not resolve differences between conditions.
Because glucose and fructose result in almost the same yield in the model, predicted final biomass of the Curated model was near 8.1~\gl{} in all conditions, whereas measured values ranged from 7.4 to 8.0~\gl{} (Supplementary Table~S5).

Maintenance and the P/O ratio could explain the remaining discrepancy on glucose.
To identify inputs that could account for this discrepancy, we changed each borrowed or derived input by $\pm$25\% (Section~\ref{sec:energy}, Supplementary Note~2).
Growth-associated maintenance moved the yield by about 5\% and the lipid fraction by about 2\% (Supplementary Fig.~S2), whereas every other input moved it by 1.2\% or less.
A lower P/O ratio had an effect of similar magnitude.
With P/O ratios of 2.25 and 2.0 instead of 2.46, the mean discrepancy fell to 3.9\% and 2.4\%.
At 2.0 the model came within 3\% of the three fermentations containing glucose but underpredicted Fru~15 by 4\%.
These lower ratios were tested only, and iJL1143 retains the template ratio of 2.46 (Section~\ref{sec:rules}).

\subsubsection{Sequential uptake and lag time improve agreement with time courses}

Both mixed fermentations used glucose before fructose (Fig.~\ref{fig:flask}b,c), an order that a growth-maximising model does not reproduce (Section~\ref{sec:dfba}).
In Glc~5 + Fru~10, for example, glucose fell to 1.5~\gl{} by 37~h while fructose stayed near 10~\gl{}.
The base setting does not impose this order, because each sugar has its own maximum uptake rate, and the simulated mixed fermentations therefore grew at 0.38~h$^{-1}$ compared with 0.20~h$^{-1}$ on a single sugar.

Adding sequential uptake and then lag time reduced the time-course RMSE several-fold (Fig.~\ref{fig:flask}g).
Sequential uptake reduced the biomass RMSE of the Curated model from 4.88~\gl{} in the base setting to 3.87~\gl{}.
Fitting the lag time then reduced the biomass RMSE to 1.16~\gl{} with one shared lag time and to 0.42~\gl{} with one lag time per condition, more than tenfold below the base setting.
Over the same steps, its sugar RMSE fell from 6.22 to 1.73~\gl{}.
The base setting reached half final biomass 16--35~h early (Supplementary Fig.~S1b), and fitting the lag time removed most of this timing error.
The timing error was $-8$ to $+7$~h with one shared lag time and below 0.2~h with one lag time per condition.
Lag time alone, however, was not sufficient.
With simultaneous uptake, the two simulated mixed fermentations still reached half final biomass 10 and 2~h early at the lowest initial biomass tested (Supplementary Note~1).

The gain from these additions lies in the time courses and not in final biomass, whereas curation improved biomass and not sugars.
The mean discrepancy in final biomass of the Curated model stayed at 6\% whether the lag time was assumed or fitted (Supplementary Note~1).
At the final setting, the Curated model still followed biomass more closely than the Draft model, with RMSEs of 0.42 and 0.75~\gl{}, whereas their sugar RMSEs differed by about 0.01~\gl{} (Fig.~\ref{fig:flask}g).
Sugar error was largest on Glc~15 and Fru~15 (Supplementary Table~S5), where measured sugar fell before online biomass rose (Fig.~\ref{fig:flask}a,d).

A fitted kinetic model of the same fermentations was closer on sugars and not on biomass \citep{vinestock2026}.
That model used 15 fitted parameters and reported mean absolute errors (MAEs) of 0.41~\gl{} for biomass and 0.57~\gl{} for substrates.
The Curated model at the final setting, with four fitted lag times, resulted in MAEs of 0.23 and 1.31~\gl{} (Section~\ref{sec:metrics}), although the two studies average replicates differently.
Because lag time is fitted to these data, this comparison shows agreement and not prediction.

\subsection{Products and medium chemistry}

\subsubsection{Transient glycerol and arabitol lie outside the model}\label{sec:res-products}

Ethanol and acetate were essentially absent.
Ethanol stayed below 0.01~\gl{} and acetate was not detected in any sample (Section~\ref{sec:hplc}).
Neither model predicted either product, in agreement with these measurements.

\begin{figure*}[tp]
  \centering
  \includegraphics[width=\textwidth,height=0.6\textheight,keepaspectratio]{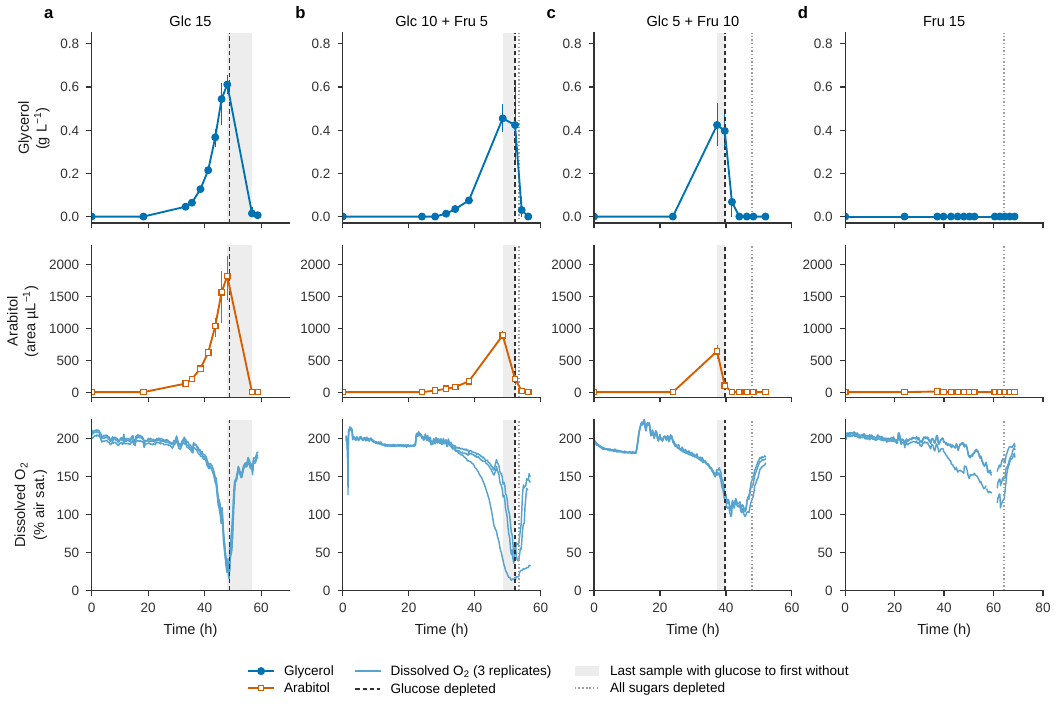}
  \caption[Transient products and dissolved oxygen]{\textbf{Glycerol and arabitol accumulate only while glucose is present.}
  Batch fermentations \citep{vinestock2026}.
  (a--d) Glycerol (top) and arabitol in area per \textmu L injected (middle), mean and range of three replicates, and dissolved \Oxy{} of the three replicates (bottom). Peak areas below zero are shown as zero. Dashed lines, mean time of glucose depletion; grey bands, last sample with glucose to first without; dotted lines, mean time at which all sugars are depleted.
  \glcfrukey}
  \label{fig:products}
\end{figure*}

Neither model predicted the transient glycerol or arabitol.
Glycerol formed only while glucose was present and peaked near glucose depletion (Fig.~\ref{fig:products}a--d).
It rose to 0.42--0.61~\gl{} in the three fermentations containing glucose but stayed below 0.01~\gl{} on Fru~15.
After the peak it was removed again, and by the last sample it had returned to 0.02~\gl{} or less.
In Glc~5 + Fru~10, it disappeared once glucose was depleted, while fructose was still being consumed.
Net removal in the first interval after glucose depletion was 0.2--0.9~\flux{} (Section~\ref{sec:energy}).
Arabitol (Section~\ref{sec:hplc}) rose together with glycerol (Fig.~\ref{fig:products}a--d) but fell faster once glucose was depleted.
In the two mixed fermentations, its mean fell by about 80\% from the last sample with glucose to the first without, whereas mean glycerol fell by less than 10\%.

Glycerol peaked near glucose depletion, whereas dissolved \Oxy{} recovered only after all sugars were depleted (Fig.~\ref{fig:products}a--c).
The two events coincided in Glc~15, lay within 5~h of each other in Glc~10 + Fru~5 and lay at least 8~h apart in Glc~5 + Fru~10.
Only Glc~5 + Fru~10 therefore separates the two events.
In this condition, glycerol accumulated while dissolved \Oxy{} stayed at about 150\% air saturation or higher, and it was removed during growth on fructose while dissolved \Oxy{} stayed near 110\%.
Glucose was depleted at about 2.2~\gl{} biomass, compared with about 6~\gl{} in Glc~15, which is consistent with the higher dissolved \Oxy{} in this condition.
Glucose depletion at lower biomass is also consistent with the lower glycerol peak in this condition.

Oxygen limitation does not appear to explain glycerol formation.
Dissolved \Oxy{} fell to 13--39\% air saturation in Glc~15 and Glc~10 + Fru~5, but it never reached zero.
In Glc~15, glycerol had already reached 89\% of its peak at 46~h, when dissolved \Oxy{} was still about 105\% air saturation (Fig.~\ref{fig:products}a).
The two single-sugar fermentations also differed in glycerol at similar oxygen.
At online biomass of 0.4--1.9~\gl{}, Glc~15 and Fru~15 grew at 0.20 and 0.16~h$^{-1}$ with dissolved \Oxy{} at 170 and 181\% air saturation.
Net glycerol formation in the same intervals was nevertheless 0.56~\flux{} on Glc~15 and zero on Fru~15.
Glycerol also did not follow dissolved \Oxy{} across the replicates of one condition.
At 48.5~h in Glc~10 + Fru~5, the three replicates held 2.0--2.5~\gl{} biomass but differed in dissolved \Oxy{}.
The replicate at 41\% air saturation held the least glycerol, 0.39~\gl{}, compared with 0.52 and 0.45~\gl{} in those at 129 and 145\%.
While glucose stayed above 2~\gl{}, the net rate had a median of 0.52~\flux{} in Glc~15 and Glc~10 + Fru~5, with no trend against dissolved \Oxy{} ($r = -0.12$, 20 intervals).
In Glc~15, it decreased to 0.03--0.23~\flux{} in the last interval before glucose depletion, when dissolved \Oxy{} was lowest.

Capping \Oxy{} uptake did not make either model form glycerol.
At caps from 80\% to 20\% of the aerobic requirement, the largest glycerol export compatible with optimal growth was zero on both sugars, whether sugar uptake was bounded or fixed at the kinetic rate (Section~\ref{sec:energy}).
The Curated model instead exported acetaldehyde and, at the lower caps, ethanol, whereas measured ethanol stayed below 0.01~\gl{}.
Acetaldehyde was not measured.
When sugar uptake was fixed and glycerol was the only carbon export, a 10\% shortfall of \Oxy{} resulted in 0.45~\flux{} of glycerol on glucose in the Curated model, compared with a measured median of 0.52~\flux{}.
The same shortfall, however, resulted in almost the same rate on fructose, 0.46~\flux{}, although no glycerol formed on Fru~15, so in the model an \Oxy{} shortfall does not distinguish the two sugars.

Part of the final biomass may form from carbon secreted earlier in the fermentation.
In Glc~15, online biomass rose from 6.1 to 7.2~\gl{} after sugars were depleted (Fig.~\ref{fig:flask}a), while glycerol and arabitol disappeared (Fig.~\ref{fig:products}a).
Online biomass is calibrated only at its end points, so this pattern is only suggestive.

\subsubsection{Medium acidification follows from growth stoichiometry}

\begin{figure*}[tp]
  \centering
  \includegraphics[width=\textwidth,height=0.6\textheight,keepaspectratio]{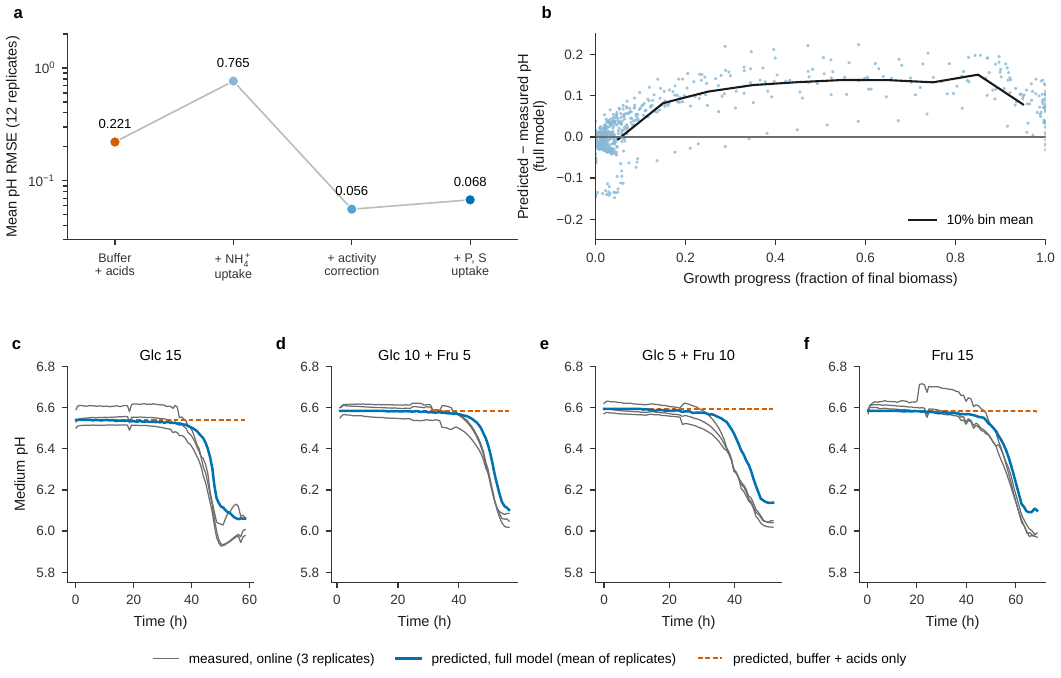}
  \caption[Medium pH in batch fermentation calculated from growth and chemistry]{\textbf{A charge balance driven by growth reproduces the fall in medium pH without fitted parameters.}
  pH computed by a charge balance driven by online biomass (Section~\ref{sec:ph}) and compared with online pH in 12 replicate fermentations \citep{vinestock2026}.
  (a) Mean pH RMSE as effects are added in an order fixed beforehand, from the buffer and measured acids alone to the full model.
  (b) Residuals (predicted minus measured) of the full model against growth progress; black line, mean in 10\% bins.
  (c--f) Measured pH (three replicates per condition) with the buffer + acids and full-model predictions.
  \glcfrukey}
  \label{fig:ph}
\end{figure*}

The fall in medium pH during growth is consistent with ammonium assimilation as the main source of acidification in this medium.
Medium pH fell from about 6.6 to about 6.0 (Fig.~\ref{fig:ph}c--f, Section~\ref{sec:ph}).
A charge balance with the buffer and measured acids alone did not reproduce this fall and predicted a nearly constant pH, with an RMSE of 0.221 (Fig.~\ref{fig:ph}a).
Adding the proton released on ammonium assimilation produced a fall, but one that was too large, with an RMSE of 0.765.
Correcting activities for the ionic strength of the medium reduced the RMSE to 0.056, and adding phosphate and sulfate uptake increased the RMSE slightly to 0.068.
No parameter was fitted to pH.

A mid-growth bias and a rebound in pH remain unexplained.
The full balance predicts pH about 0.13 units above the measurement in mid-growth (Fig.~\ref{fig:ph}b).
It also does not reproduce the rebound in pH after sugar depletion in Glc~15 (Fig.~\ref{fig:ph}c).
Because the balance uses online biomass, it tests the chemistry and is not a forward prediction.

\section{Discussion}

Our hypothesis held for both how much biomass forms and when it forms, the two quantities that a model for mycoprotein process design must estimate.
Rule-based curation alone brought predicted final biomass closer to measurement in oxygen-enriched batch fermentation without fitted parameters, and sequential uptake and lag time brought the biomass time courses into close agreement.
Lag time was fitted to the same time courses, so the timing result shows agreement rather than prediction.
The informative result is that a fitted lag time alone could not align the mixed fermentations without sequential uptake.
The advantage over the Draft model lies in biomass, and both models described sugars equally well.
The main remaining discrepancy is an overprediction of final biomass by 6--9\% in the fermentations containing glucose.

These results apply to oxygen-enriched batch fermentation on glucose and fructose at one sugar level.
Sequential uptake of glucose and fructose is consistent with carbon catabolite repression in filamentous fungi \citep{adnan2018}.
A kinetic model of the same fermentations also assumed that glucose is preferred \citep{vinestock2026}.
That model described sugars more closely than the Curated model and biomass less closely, with 15 fitted parameters compared with four.
The two approaches are complementary, because the kinetic model describes sugar consumption better and the Curated model estimates yield before any data are fitted.

The directionality and carrier rules, and not energy-generating cycles, set the yield here.
Energy-generating cycles are common in uncurated models and typically inflate maximal biomass production by 25\% \citep{fritzemeier2017}.
In this study, the cycle tests did not determine the yield.
Models that failed them retained the yield of the Curated model when the directionality and carrier rules were in place, and the Curated model without the carrier rule passed every test with an ATP yield above the theoretical maximum.
The role of directionality fits the view of that study that such errors can reflect wrong assumptions about reversibility.
The carrier direction matters because mitochondrial ATP could otherwise reach the cytosol as GTP, without the proton cost of the ADP/ATP carrier.
Directionality and transport errors can inflate ATP yield in this way, and curated models of \textit{Escherichia coli}, yeast and human metabolism are therefore checked against realistic ATP yields \citep{fleming2009,lu2019,swainston2016}.
With this cost in place the P/O ratio lies close to measured values \citep{hinkle2005} and to the ratios set in iMA871 \citep{andersen2008}, the source of the growth-associated maintenance.
Maximum ATP yield should therefore be tested alongside energy-generating cycles when a model is transferred from a related species.

A lower P/O ratio and unmeasured products may each explain part of the remaining discrepancy on glucose.
Products remaining at the end of the fermentation cannot explain it, because glycerol and arabitol concentrations return to about zero.
A lower P/O ratio could contribute, because effective P/O ratios below the mechanistic value have been reported for the filamentous fungus \textit{Penicillium roqueforti} in solid-substrate culture \citep{larroche1992}, and a ratio of 2.0 brought the three fermentations containing glucose within 3\%.
The same ratio underpredicted Fru~15, so a single P/O ratio cannot explain both sugars. Yield data alone also cannot separate the P/O ratio from growth-associated maintenance \citep{vonkamp2023}.
Unmeasured products may also contribute, because medium phosphate masks pyruvic and citric acids (Section~\ref{sec:hplc}).
The mid-growth bias in pH and the rebound after sugar depletion are both consistent with an unmeasured acid that forms during growth and is consumed once sugars are depleted.
Polyols other than these two are a further candidate, because
deleting a mannitol-1-phosphate dehydrogenase in another \Fven{} strain decreased supernatant mannitol and increased biomass by 9\% \citep{liu2026}.
A growth-maximising model secretes a by-product only when a constraint makes its secretion part of the optimal solution, as the capped \Oxy{} uptake did for acetaldehyde and ethanol (Section~\ref{sec:res-products}).
Without oxygen or nitrogen limitation, carbon secreted as a by-product is unavailable for growth, so the optimal solution contains no by-product unless a further constraint makes its secretion optimal.
Glycerol formed at high dissolved \Oxy{} and not on fructose at similar growth, which argues against oxygen limitation and suggests a regulatory effect that a growth-maximising model does not contain.
Because removal after glucose depletion was about as fast as formation, its absence on fructose could also reflect immediate reuse.

These conclusions are limited by their scope, inputs and measurements.
They rest on one strain, four conditions, two sugars and one sugar level, with three replicates each.
Within that scope, four inputs could bias the comparison.
Maintenance is borrowed, cell wall, lipid, DNA and the small-molecule pool are not measured in this strain, lag time is fitted to the same data, and initial sugars are nominal, with sugar concentrations scaled to them.
The measurements also have limits.
CO$_2$ was not measured, so the carbon balance cannot be closed, and online biomass is calibrated only at its end points.
The yield comparison nevertheless rests on the gravimetric final dry weight, which does not depend on the online calibration.
That dry weight includes ash, which the model lacks, so both discrepancies could be up to about 6 percentage points larger without changing their order (Sections~\ref{sec:metrics} and \ref{sec:res-biomass}).

\section{Conclusions}

iJL1143, the Curated model of \Fven{} A3/5, predicts final biomass in oxygen-enriched batch fermentation with a mean discrepancy of 6\% without fitted parameters, compared with 20\% for the Draft model.
Sequential sugar uptake and a fitted lag time reduce its biomass time-course RMSE by more than tenfold.
In oxygen-enriched batch fermentation, it can therefore estimate the biomass obtainable from a sugar feedstock and describe the course of a batch fermentation.
Groups building a model for a related fungus should apply thermodynamic directionality rules, check the proton cost of mitochondrial ATP export, and test maximum ATP yield as well as energy-generating cycles before any comparison with data.
They should also add sequential sugar uptake and lag time before judging time courses.

\section*{CRediT authorship contribution statement}
J.L.: Conceptualization, Methodology, Software, Data curation, Formal analysis, Investigation, Validation, Visualization, Writing -- original draft, Writing -- review \& editing.
T.V.: Conceptualization, Data curation, Formal analysis, Investigation, Writing -- review \& editing.
M.G.: Conceptualization, Funding acquisition, Supervision, Writing -- review \& editing.

\section*{Declaration of competing interest}
T.V.\ and M.G.\ have received research funding from Quorn, which produces mycoprotein from \Fven{}.
J.L.\ declares no competing interests.

\section*{Data availability}
The batch fermentation data were published previously \citep{vinestock2026}.
The Draft and Curated (iJL1143) models in SBML format and the curation change log are available at \url{https://github.com/MGuo-Lab/Fvenenatum-GEM}.
The time-course measurements of the batch fermentations and the result files underlying every figure and table are available in the same repository.
The \Fven{} A3/5 proteome is available from UniProt (UP000245910), and its genome is available from the European Nucleotide Archive (PRJEB7533) \citep{king2018}.
The template model iJB1325 is available with its original publication \citep{brandl2018}.

\section*{Code availability}
The code for model reconstruction, curation, simulation, analyses and figures is available at \url{https://github.com/MGuo-Lab/Fvenenatum-GEM}.
Supplementary Table~S6 lists the software used, with the version and role of each package.

\section*{Declaration of generative AI and AI-assisted technologies in the manuscript preparation process}
During the preparation of this work the authors used Claude (Anthropic) to revise the text for clarity and flow and to check reported values against the analysis outputs.
After using this tool, the authors reviewed and edited the content as needed and take full responsibility for the content of the published article.

\section*{Acknowledgements}
J.L.\ and M.G.\ would like to acknowledge financial support from the King's College London Net Zero Centre Ph.D.\ Scholarship scheme.
T.V.\ and M.G.\ acknowledge funding from Quorn and the BBSRC under project reference 2725958.
The authors acknowledge the use of resources provided by the Isambard-AI National AI Research Resource (AIRR) \citep{mcintoshsmith2024}.
Isambard-AI is operated by the University of Bristol and is funded by the UK Government's Department for Business, Innovation, Science and Trade (DBIST) via UK Research and Innovation; and the Science and Technology Facilities Council [ST/AIRR/I-A-I/1023].

\bibliographystyle{elsarticle-num}
\bibliography{references}

\clearpage
\onecolumn
\section*{Supplementary material}
\renewcommand{\topfraction}{0.98}
\renewcommand{\bottomfraction}{0.98}
\renewcommand{\textfraction}{0.02}
\renewcommand{\floatpagefraction}{0.4}
\setcounter{topnumber}{6}
\setcounter{totalnumber}{6}
\makeatletter
\setlength{\@fptop}{0pt}
\setlength{\@fpsep}{16pt}
\setlength{\@fpbot}{0pt plus 1fil}
\makeatother
\setlength{\intextsep}{14pt plus 2pt minus 2pt}
\setlength{\textfloatsep}{10pt plus 2pt minus 2pt}
\setlength{\floatsep}{12pt plus 2pt minus 2pt}

\setcounter{figure}{0}
\setcounter{table}{0}
\renewcommand{\thefigure}{S\arabic{figure}}
\renewcommand{\thetable}{S\arabic{table}}
\renewcommand{\theHfigure}{S\arabic{figure}}
\renewcommand{\theHtable}{S\arabic{table}}
\renewcommand{\figurename}{Supplementary Figure}
\renewcommand{\tablename}{Supplementary Table}

\begin{table}[!ht]
  \centering
  \caption[Curation rules and their effect]{\textbf{Curation rules and their effect.}}
  \label{tab:rules}
  \small
  \begin{tabularx}{\textwidth}{@{}c >{\raggedright\arraybackslash}X >{\raggedright\arraybackslash}X >{\raggedright\arraybackslash}p{3.4cm}@{}}
    \toprule
    Stage & Rule & Basis & Effect \\
    \midrule
    1 & Orthology filter: reactions kept only with orthology support (tier 1 RBH, tier 2 orthogroup, complexes kept when one subunit maps) & Orthology & 85 template reactions removed \\
    2 & Reaction recovery by EC number and gap-filling for biomass synthesis & UniProt EC of the \Fven{} proteome; tasks of Thiele and Palsson~\citep{thiele2010} & +6 by EC number, +3 by gap-filling \\
    3 & Secretion restricted to a gene-supported transporter or literature support & List of metabolites with citations & 330 exchange reactions closed \\
    4 & Redox rules against cost-free NADH--NADPH and cytosol--mitochondria routes, and NAD(P) reaction directionality & eQuilibrator $|\ln\Gamma| \geq 7$ \citep{noor2012} & 15 redox actions (3 reactions removed), 20 directionality changes \\
    5 & Biomass normalised to the composition of growing mycelium (Supplementary Table~\ref{tab:biomass}), NGAM 1.9 & \citep{towersey1977,fda2002grn91,andersen2008} & 1~g\,mmol$^{-1}$ \\
    6 & Thermodynamic directionality of all metabolic reactions & eQuilibrator $|\ln\Gamma| \geq 7$ \citep{noor2012} & 29 directionality changes (incl.\ phosphoketolase) \\
    6 & Organic-acid export pays 1 ATP per negative charge & \citep{vanmaris2004,vanmaris2004aem,abbott2009} & 22 transports made import-only, 22 ATP-coupled export reactions added \\
    6 & Mitochondrial GTP/GDP carrier imports GTP & \citep{vozza2004} & 1 reaction \\
    \bottomrule
  \end{tabularx}

  \smallskip
  \parbox{\textwidth}{\footnotesize Every rule is applied to all reactions it covers. The curation change log, available with the model (Data availability), lists each reaction with its evidence.}
\end{table}
\begin{table}[!ht]
  \centering
  \caption[Biomass composition]{\textbf{Biomass composition of growing \Fven{} A3/5 mycelium, used in both models.}}
  \label{tab:biomass}
  \small
  \begin{tabularx}{\textwidth}{@{}>{\raggedright\arraybackslash}p{2.6cm} c c >{\raggedright\arraybackslash}X >{\raggedright\arraybackslash}p{3.0cm}@{}}
    \toprule
    Fraction & Dry mass (\%) & Model (\%) & Basis & Source \\
    \midrule
    Protein & 37 & 38.9 & Measured in A3/5 mycelium before RNA reduction (37.3\%), continuous culture on hydrolysed starch at 0.14~h$^{-1}$ & \citep{towersey1977} \\
    RNA & 9 & 9.5 & Measured in untreated A3/5 mycelium, 7--12\% in the same culture and 8--9\% at 0.19~h$^{-1}$ & \citep{towersey1977,trinci1992} \\
    Cell wall & 17 & 17.9 & Cell-wall material of mycoprotein (22--28\%) times the 0.68~g of product left per gram of mycelium after RNA reduction & \citep{fda2002grn91,towersey1977} \\
    Lipid & 9 & 9.5 & Fat of mycoprotein (12--14\%) times 0.68 & \citep{fda2002grn91,towersey1977} \\
    DNA & 0.9 & 0.9 & \textit{Aspergillus oryzae}, glucose-limited chemostat (0.8--1.0\%) & \citep{pedersen1999} \\
    Ash & 5 & -- & Batch mycelium (5.3\%); mycoprotein specification of 5\% or less & \citep{solomons1976,fda2002grn91} \\
    Small-molecule pool & 22 & 23.3 & Remainder & -- \\
    \bottomrule
  \end{tabularx}

  \smallskip
  \parbox{\textwidth}{\footnotesize Model, fraction of ash-free dry mass in the biomass reaction of both models, which weighs 1~g\,mmol$^{-1}$. Cell wall and lipid assume that both are fully retained during RNA reduction.}
\end{table}
\begin{table}[!ht]
  \centering
  \caption[Effect of each curation stage]{\textbf{Effect of each curation stage on energetics and yield.}}
  \label{tab:stages}
  \small
  \begin{tabularx}{\textwidth}{@{}X *{6}{>{\centering\arraybackslash}p{1.25cm}}@{}}
    \toprule
     & \multicolumn{3}{c}{Curated without the stage} & \multicolumn{3}{c}{Draft with the stage only} \\
    Stage & ATP & Tests failed & Yield & ATP & Tests failed & Yield \\
    \midrule
    1 Orthology filter & 29.9 & 2 & 0.523 & 49.1 & 2 & 0.599 \\
    2 Recovery by EC number & 29.9 & 0 & 0.522 & 49.1 & 2 & 0.599 \\
    2 Gap-filling & 29.9 & 0 & 0.522 & 49.1 & 2 & 0.599 \\
    3 Secretion restriction & 29.9 & 0 & 0.522 & 49.1 & 2 & 0.599 \\
    4 Redox rules & 29.9 & 2 & 0.523 & 49.1 & 2 & 0.599 \\
    4 NAD(P) directionality & 30.0 & 2 & 0.525 & 37.1 & 2 & 0.559 \\
    6 Directionality, all reactions & 29.9 & 0 & 0.523 & 37.1 & 2 & 0.561 \\
    6 Organic-acid export cost & 29.9 & 0 & 0.522 & 49.1 & 2 & 0.599 \\
    6 GTP/GDP carrier direction & 37.1 & 0 & 0.559 & 49.1 & 2 & 0.599 \\
    4 and 6 directionality together & 49.1 & 2 & 0.599 & & & \\
    4 directionality and carrier & & & & 29.9 & 2 & 0.523 \\
    6 directionality and carrier & & & & 30.0 & 2 & 0.525 \\
    \midrule
    Reference: Curated & 29.9 & 0 & 0.522 & & & \\
    Reference: Draft & & & & 49.1 & 2 & 0.599 \\
    \bottomrule
  \end{tabularx}

  \smallskip
  \parbox{\textwidth}{\footnotesize ATP, maximum mol ATP per mol glucose. Tests failed, energy-generating cycle tests. Yield, aerobic biomass yield on glucose (g\,g$^{-1}$) at an uptake of 1~\flux. Stage~5 sets biomass and maintenance, which both models share. Stage~1 applied to the Draft model retains the nine reactions that Stage~2 restores.}
\end{table}
\begin{table}[!ht]
  \centering
  \caption[MEMOTE scores]{\textbf{MEMOTE scores of the two models (version 0.17).}}
  \label{tab:memote}
  \small
  \begin{tabular}{@{}lcc@{}}
    \toprule
    Section & Draft & Curated \\
    \midrule
    Total score & 29\% & 29\% \\
    Model consistency & 54\% & 55\% \\
    Metabolite annotation & 62\% & 62\% \\
    Reaction annotation & 48\% & 45\% \\
    Gene annotation & 0\% & 0\% \\
    SBO terms & 0\% & 0\% \\
    \bottomrule
  \end{tabular}
\end{table}

\begin{table}[!ht]
  \centering
  \caption[Measured and predicted values per batch fermentation]{\textbf{Measured and predicted values per batch fermentation.}}
  \label{tab:flask-values}
  \footnotesize
  \setlength{\tabcolsep}{4pt}
  \begin{tabularx}{\textwidth}{@{}X c c *{4}{c} c c c@{}}
    \toprule
     & \multicolumn{2}{c}{Measured} & \multicolumn{2}{c}{Curated} & \multicolumn{2}{c}{Draft} & \multicolumn{3}{c}{Curated, lag time per condition} \\
    Condition & Biomass & Yield & Biomass & Discrep. & Biomass & Discrep. & $x_0$ & RMSE$_X$ & RMSE$_S$ \\
    \midrule
    Glc~15 & 7.44 $\pm$ 0.24 & 0.50 & 8.11 & +8.9 & 9.16 & +23.0 & $3.9\times10^{-4}$ & 0.48 & 3.04 \\
    Glc~10 + Fru~5 & 7.67 $\pm$ 0.38 & 0.51 & 8.10 & +5.6 & 9.14 & +19.2 & $2.0\times10^{-4}$ & 0.28 & 0.90 \\
    Glc~5 + Fru~10 & 7.48 $\pm$ 0.10 & 0.50 & 8.11 & +8.4 & 9.15 & +22.3 & $1.1\times10^{-3}$ & 0.57 & 0.87 \\
    Fru~15 & 7.98 $\pm$ 0.34 & 0.53 & 8.12 & +1.7 & 9.16 & +14.8 & $5.5\times10^{-5}$ & 0.36 & 2.11 \\
    \bottomrule
  \end{tabularx}

  \smallskip
  \parbox{\textwidth}{\footnotesize Biomass, $x_0$ (fitted initial biomass) and RMSE in g\,L$^{-1}$; yield in g\,g$^{-1}$; discrepancy (Discrep.) in \% of the measured mean, positive for overprediction. Measured values, from the fermentations of Vinestock and Guo~\citep{vinestock2026}, are mean $\pm$ s.d., $n=3$. Predicted biomass and discrepancy use sequential uptake and nothing fitted. RMSE$_X$ biomass, RMSE$_S$ sugars.}
\end{table}

\begin{table}[!ht]
  \centering
  \caption[Software used]{\textbf{Software used for reconstruction, curation, simulation and analysis.}}
  \label{tab:software}
  \small
  \begin{tabularx}{\textwidth}{@{}>{\raggedright\arraybackslash}p{2.7cm} >{\raggedright\arraybackslash}p{1.5cm} >{\raggedright\arraybackslash}X >{\raggedright\arraybackslash}p{3.1cm}@{}}
    \toprule
    Software & Version & Role in this study & Reference \\
    \midrule
    BLAST+ & 2.16.0 & Reciprocal best hits between the template and \Fven{} proteomes with blastp, keeping five target sequences per query & \citep{camacho2009} \\
    OrthoFinder & 2.5.5 & Orthogroups across 10 proteomes: \textit{Aspergillus fumigatus}, \textit{A.~nidulans}, \textit{A.~niger} CBS 513.88, \textit{A.~oryzae}, \textit{A.~terreus}, \textit{Fusarium fujikuroi}, \textit{F.~graminearum}, \textit{F.~venenatum}, \textit{F.~verticillioides} and \textit{Trichoderma reesei} & \citep{emms2019} \\
    DIAMOND & 2.2.8 & Sequence aligner inside OrthoFinder & \citep{buchfink2021} \\
    Biopython & 1.88 & Smith--Waterman alignment of complex subunits without an ortholog against the whole proteome & \citep{cock2009} \\
    eQuilibrator API & 0.8.1 & Reversibility index for the directionality rules & \citep{noor2012,beber2022} \\
    COBRApy & 0.32.1 & Model input and output, flux balance analysis and each dFBA step & \citep{ebrahim2013} \\
    GLPK & 5.0 & Linear programming solver called through COBRApy & \url{https://www.gnu.org/software/glpk/} \\
    MEMOTE & 0.17.0 & Model quality scores & \citep{lieven2020} \\
    SciPy & 1.18.1 & Root finding and bounded optimisation for the lag time and the yield sensitivity & \citep{virtanen2020} \\
    Python & 3.12.3 & Scripting language; the orthology step used 3.11.18 & -- \\
    NumPy, pandas, Matplotlib & 2.5.3, 2.3.3, 3.11.2 & Data handling and figures & -- \\
    \bottomrule
  \end{tabularx}
\end{table}

\FloatBarrier
\subsection*{Supplementary Note 1. Lag time in batch fermentation}

\begin{figure}[!ht]
  \centering
  \includegraphics[width=\textwidth,height=0.6\textheight,keepaspectratio]{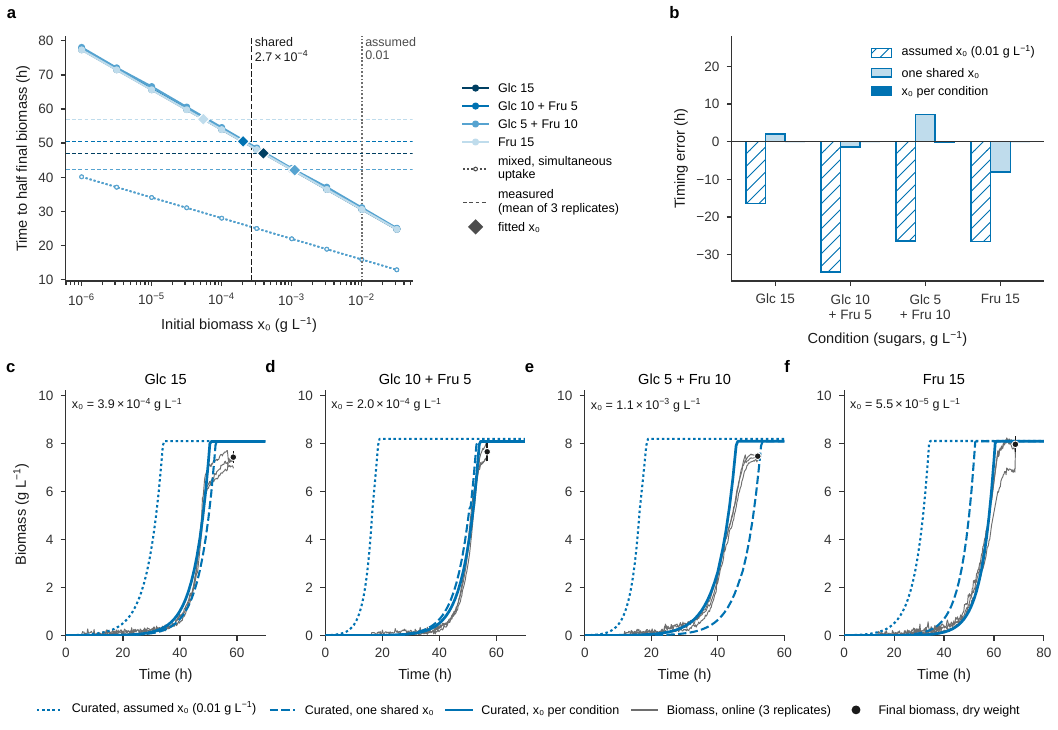}
  \caption[Initial biomass fitted to the timing of growth]{\textbf{Initial biomass fitted to the timing of growth.}
  Curated model and batch fermentations \citep{vinestock2026}.
  (a) Time to half final biomass against initial biomass $x_0$; dashed lines, measured mean of three replicates; diamonds, fitted $x_0$. The dotted line shows the mixed fermentations with simultaneous uptake.
  (b) Timing error with the assumed $x_0$ (0.01~\gl{}) and simultaneous uptake, as in the base setting, one shared $x_0$ and one $x_0$ per condition.
  (c--f) Biomass over time for the three settings, with online biomass of the three replicates (grey) and final dry weight (black point, mean $\pm$ s.d.).
  \glcfrukey}
  \label{fig:lag}
\end{figure}

Fitted initial biomass sets the timing of growth and leaves final biomass nearly unchanged.
Initial biomass was fitted to the time to half final biomass of each condition, with sequential uptake in the mixed fermentations (Section~2.2.1, Supplementary Fig.~\ref{fig:lag}).
Fitted values for the Curated model ranged from $5.5\times10^{-5}$ to $1.1\times10^{-3}$~\gl{}, and one shared value was $2.7\times10^{-4}$~\gl{}.
The mean discrepancy in final biomass was 6.2\% at the assumed value and 6.1\% when fitted.

\subsection*{Supplementary Note 2. Sensitivity of the yield}

Yield depends mainly on maintenance and the P/O ratio, and little on the assumed sugar $K_\mathrm{m}$.
Settings are given in Section~2.2.2, and the change for each input in Supplementary Fig.~\ref{fig:sensitivity}.
The Curated model predicts 0.54~g\,g$^{-1}$ on both sugars, close to the measured yield of 0.53~g\,g$^{-1}$ on fructose.
Reaching the measured yield on glucose would require growth-associated maintenance to be increased 1.4-fold or non-growth-associated maintenance 3.7-fold.
Reducing the P/O ratio for NADH from 2.46 to 2.25 and 2.0 results in a maximum of 27.5 and 24.7~mol ATP per mol glucose.
The yields are then 0.53 and 0.51~g\,g$^{-1}$.
Across a $K_\mathrm{m}$ of 0.025--4.5~mM, final biomass changed by less than 1\%.
With initial biomass held at its fitted value, the time to half final biomass moved by less than 0.3~h at 0.025~mM and by up to 9~h at 4.5~mM.

\begin{figure}[!ht]
  \centering
  \includegraphics[width=\textwidth,height=0.45\textheight,keepaspectratio]{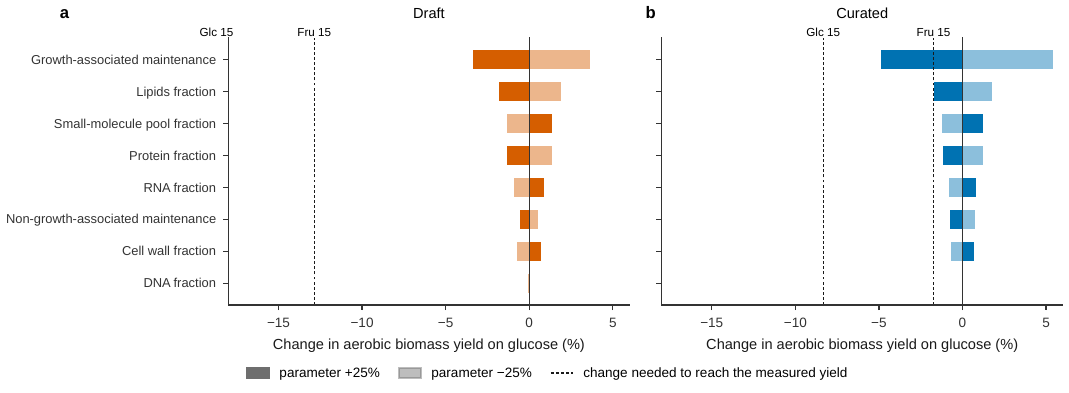}
  \caption[Sensitivity of the aerobic biomass yield]{\textbf{Sensitivity of the aerobic biomass yield to biomass composition and maintenance.}
  (a) Draft model and (b) Curated model. Change in yield on glucose for each parameter changed by $\pm$25\% (Section~2.2.2). Dashed lines, the change that would reach the measured yield of Glc~15 and Fru~15.}
  \label{fig:sensitivity}
\end{figure}

\enlargethispage{\baselineskip}
\subsection*{Supplementary Note 3. Identity of the 16.1~min peak}

\begin{figure}[!ht]
  \centering
  \includegraphics[width=0.8\textwidth]{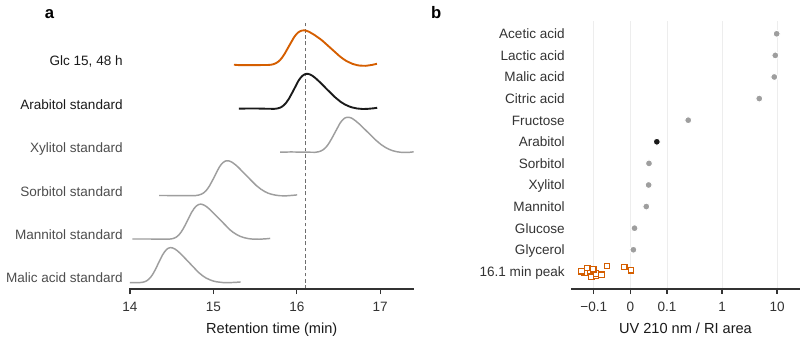}
  \caption[The 16.1~min peak matches arabitol]{\textbf{The 16.1~min peak matches an arabitol standard and is not a carboxylic acid.}
  (a) Refractive-index signal near 16~min for Glc~15 at 48~h and for single standards at 1~\gl{}, each scaled to its own peak. Standard retention times are aligned to the fermentation run with glucose and glycerol. Dashed line, apex of the fermentation peak. (b) Ratio of absorbance at 210~nm to refractive-index area, in \textmu AU\,s per \textmu V\,s, for standards and for the 11 Glc~15 samples with a 16.1~min peak area above 2000~\textmu V\,s.}
  \label{fig:peak16}
\end{figure}

The 16.1~min peak matches an arabitol standard in retention time.
Arabitol, xylitol, sorbitol, mannitol and malic acid standards at 1~\gl{} were injected on the same instrument and column, with the same mobile phase, flow and column temperature, in a sequence run before the fermentations.
Retention times of that sequence were aligned to each fermentation run with glucose and glycerol as reference peaks.
An alignment of this kind, tested on arabitol between two standard sequences, erred by 0.04~min.
In Glc~15 the peak eluted at 16.10~min, 0.03~min before the aligned arabitol standard (Supplementary Fig.~\ref{fig:peak16}a).
Xylitol, sorbitol, mannitol and malic acid eluted at least 0.5~min away.
No arabinose standard was run at 30\,\textdegree C.
At 25\,\textdegree C on the same instrument, arabinose eluted at 0.75 of the interval from glucose to xylitol, against 0.85 for arabitol.
At 30\,\textdegree C this places arabinose 0.28~min before the 16.1~min peak.

UV absorbance excludes a carboxylic acid but does not separate polyols from one another.
The ratio of absorbance at 210~nm to refractive-index area was between $-0.17$ and 0.00 for the 16.1~min peak in Glc~15 (Supplementary Fig.~\ref{fig:peak16}b).
The ratio was 4.8--9.8 for citric, malic, lactic and acetic acid standards and 0.04--0.07 for the four polyol standards.
The peak appeared only in fermentations with glucose and was absent from all 49 injections of Fru~15.
Co-injection with arabitol or mass spectrometry would be needed to confirm the assignment, and the enantiomer is not resolved.

\end{document}